\documentclass[11pt,twoside]{article}
\usepackage{asp2004}
\usepackage{psfig}
\usepackage{epsf}
\usepackage{graphics}
\usepackage{lscape}
\def\edcomment#1{\iffalse\marginpar{\raggedright\sl#1\/}\else\relax\fi}
\reversemarginpar

\begin{document}
\title{SOAP and the Interstellar Froth}
\author{R. T\"ullmann}
\affil{Astronomisches Institut, Ruhr-Universit\"at Bochum, Universit\"atsstr. 150, D-44780 Bochum, Germany}
\author{M. R. Rosa} 
\affil{Space Telescope European Coordinating Facility, c/o European Southern Observatory, Karl-Schwarzschild-Str. 2, D-85748 Garching, Germany}
\author{R.--J. Dettmar}
\affil{Astronomisches Institut, Ruhr-Universit\"at Bochum, Universit\"atsstr. 150, D-44780 Bochum, Germany}

\begin{abstract}
We investigate, whether the alleged failure of standard photoionization codes of the Diffuse Ionized Gas (DIG) is simply caused by geometrical effects and the insufficient treatment of the radiative transfer.  
Standard photoionization models are applicable only to homogeneous and spherical symmetric nebulae with central ionizing stars, whereas the geometry of disk galaxies requires a 3D distribution of ionizing sources in the disk which illuminate the halo. This change in geometry together with a proper radiative transfer model is expected to substantially influence ionization conditions.  
Therefore, we developed a new and sophisticated 3D Monte Carlo photoionization code, called {\it SOAP} (Simulations Of Astrophysical Plasmas), by adapting an existing 1D code for HII-regions \citep*{och} such, that it self-consistently models a 3D disk galaxy with a gaseous DIG halo.  
First results from a simple (dust-free) model with exponentially decreasing gas densities are presented and the predicted ionization structure of disk and halo is discussed.  Theoretical line ratios agree well with observed ones, e.g,. for the halo of NGC\,891.
Moreover, the fraction of ionizing photons leaving the halo of the galaxy is plotted as a function of varying gas densities. This quantity will be of particular importance for forthcoming studies, because rough estimates indicate that about 7\% ionizing photons escape from the halo and contribute to the ionization of the IGM.
Given the relatively large number density of normal spiral galaxies, OB-stars could have a much stronger impact on the ionization of the IGM than AGN or QSOs. 
\end{abstract}
\thispagestyle{plain}

\section{Introduction}
Various observational and theoretical studies have been carried out which aimed at the investigation of possible ionization mechanisms of the Diffuse Ionized Gas (DIG, also called Interstellar Froth \citep{hunter90}) in the Milky Way \citep[e.g.,][]{rey84,rey99,wood04a,wood04b} as well as in external galaxies \citep[e.g.,][]{rand97,tu00a,tu00b,coll01,otte02,mil03,tu05}. As a main result of these studies, photoionization by OB-stars turned out to be the main ionization mechanism of this gas. It soon became obvious that existing photoionization models cannot predict the observed DIG spectrum correctly. In order to reconcile theory with observations, the current strategy is to include additional heating and ionization mechanisms, such as shock ionization, turbulence or magnetic reconnection (to name only a few).
Even this didn't solve the inconsistencies completely.

Therefore, we decided to analyze the existing models critically according to the radiative transfer and the geometry they consider. The most commonly used codes to model DIG are CLOUDY \citep{ferland98} and the one written by Mathis \citep{ma86} with its two modifications presented by \citet{dom} and \citet{ma00}. In addition, an implementation of Ferland's CLOUDY has been published by \citet{soko} for the edge-on galaxy NGC\,891.

There are two severe shortcomings common to all these codes: (1) the problem of the radiative transfer is solved analytically which requires unrealistic approximations (OTS, OWO, etc.) and underestimate the {\it diffuse} ionizing radiation field. (2) the geometry assumed by these codes is spherically symmetric which is not realistic to model galaxies with three dimensional stellar disks and extended gaseous halos. Hence, photoionization by OB-stars as the main heat source of the DIG should not be abandoned, unless realistic pure photoionization models have proven to be unsuccessful in reproducing the observed line ratios.

\section{Simulations Of Astrophysical Plasmas ({\it SOAP})}
Since the radiative transfer can only be solved correctly by means of statistical methods, a new self-consistent 3D Monte Carlo (MC) code, called {\it SOAP} (Simulations Of Astrophysical Plasmas), has been developed. MC techniques are a proven means of calculating the radiative transfer in an inhomogeneous medium. As an example, \citet{abbott} and \citet{lucy} developed a self-consistent MC code which successfully models the multi-line transfer in the wind of Wolf-Rayet and other early-type stars.  This model was modified by \citet{och} in order to self-consistently simulate spherical steady state gaseous nebulae with arbitrary densities. 
A comparison of their results with those produced by independent analytical photoionization codes for the standard {\it Meudon H${\scriptstyle I\hspace{-0.04cm}I}$-region} \citep{ferland95} evidenced the correctness of the MC model code.  

Based on this work, we extended the 1D model grid to 3D, introduced a 3D stellar distribution, assigned individual gas densities to every grid point, and adapted the radiative transfer, correspondingly.  As a result, new comprehensive data, such as the ionization structure of H, He, and heavier ions at high distances above the disk plane as well as first line ratios, could be obtained. 
We are aware that the interaction of the radiation field with dust particles is currently neglected. However, contributions from dust will be modeled as soon as the effects of pure photoionization for different parameter settings are investigated and can be disentangled from a dusty plasma.

Before the modeling of gaseous galaxy halos was started, we made sure that the standard {\it Meudon H${\scriptstyle I\hspace{-0.04cm}I}$-region} results obtained with the Och-model could be fully reproduced by {\it SOAP}. A detailed description of the code is given in \citet{tu02}.

\subsection{The model grid}
Since we want to investigate the ionization conditions mainly along
the $z$ direction, Cartesian coordinates are used rather than
spherical or cylindrical ones. Our model galaxy extends from the
center 15\,kpc in each direction. The disk is assumed to have a radial
extend of 9\,kpc and a height of 1.5\,kpc.  The whole geometry is
subdivided into 1400 cells, with 10 cells in $x$ and $y$-direction,
respectively, and 14 cells along the $z$-axis.  Except for the disk
region all cells have constant lengths of 1.5\,kpc and identical volumes.  
Within the disk, there are 5 vertical layers with heights of 100, 100, 
200, 400, and 700\,pc. Stars are assumed to be present only in a thin disk which 
occupy the innermost 100\,pc. 
The remaining layers are filled up with gas of pre-selected density and metal 
abundance. Above the disk, gas densities are significantly lower and drop
exponentially as a function of $z$, the distance above the disk.

Although the selected geometry underestimates the radial dimensions of a ``real''
spiral galaxy by a factor of 2 and is only sensitive to changes of the
radiation field on scales of 1.5\,kpc the chosen setup represents the best
compromise between spatial resolution and computing time.

\subsection{A simple model}
{\it SOAP} has been developed to primarily investigate DIG properties in
edge-on galaxies. This would be similar to spectroscopic observations where 
the slit is aligned along the minor axis of a galaxy. Here, a typical offset of 
$x=6.0$\,kpc from the galactic center was assumed. At this position, all relevant 
emission coefficients from recombination and forbidden lines are summed up and 
volume averaged along the $y$-axis for each of the 14 $z$-layers, separately.

A model halo with exponentially decreasing densities is simulated assuming the 
relative elemental abundances of the {\it Meudon H${\scriptstyle I\hspace{-0.04cm}I}$-region} (H\,=\,1.0, He\,=\,0.1, C\,=\,2.2$\times 10^{-4}$, N\,=\,4.0$\times 10^{-5}$, O\,=\,3.3$\times 10^{-4}$, Ne\,=\,5.0$\times 10^{-5}$, and S\,=\,9.0$\times 10^{-6}$). Densities in the lowermost layer start at $n_{gas}=3.0\,{cm}^{-3}$ and drop to $2.5\times 10^{-4}$\,cm$^{-3}$ at $z=15$\,kpc. Given the symmetry of the problem, it is sufficient to model only one octant of the galaxy. Within the volume occupied by the disk, approximately 640 O6 stars (blackbody) are distributed in a thin (100\,pc) stellar layer with a radius of 9.0\,kpc. 

\subsection{Results and discussion}
A typical model run consists of $\sim$\,150 iterations with 10$^{7}$ photons
per iteration. We made sure that the solution remained stable even after 450 model 
runs.

\begin{figure}[!t]
\plotone{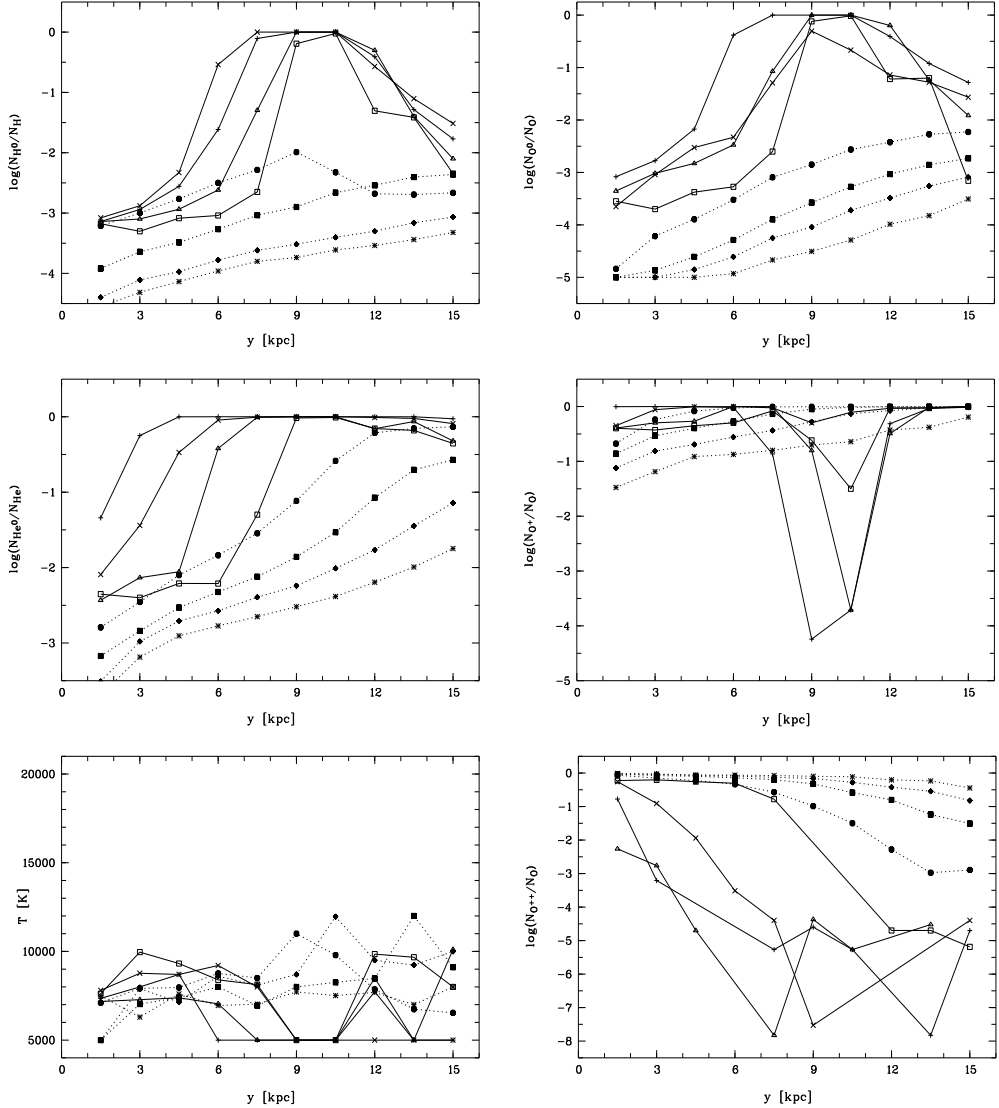}
\caption{Ionization fractions for a halo of inhomogeneous gas density. Plotted are fractions for the different oxygen ionization stages (panels to the right), for H$^{0}$ (upper left) and He$^{0}$ (middle left), together with electron temperatures (lower left) as a function of $y$ for various $z$-heights.  Solid lines address ionization fractions for the disk at distances of 0.1 (open squares), 0.4 (open triangles), 0.8 (``+''-signs), and 1.5\,kpc (``x''-signs) from the plane.  Dashed lines represent the corresponding fractions for the halo at 4.5 (filled hexagons), 9.0 (filled squares), 13.5 (filled lozenges), and 15.0\,kpc (asterisks).}
\label{f1}
\end{figure}

\begin{figure}[!t]
\plotone{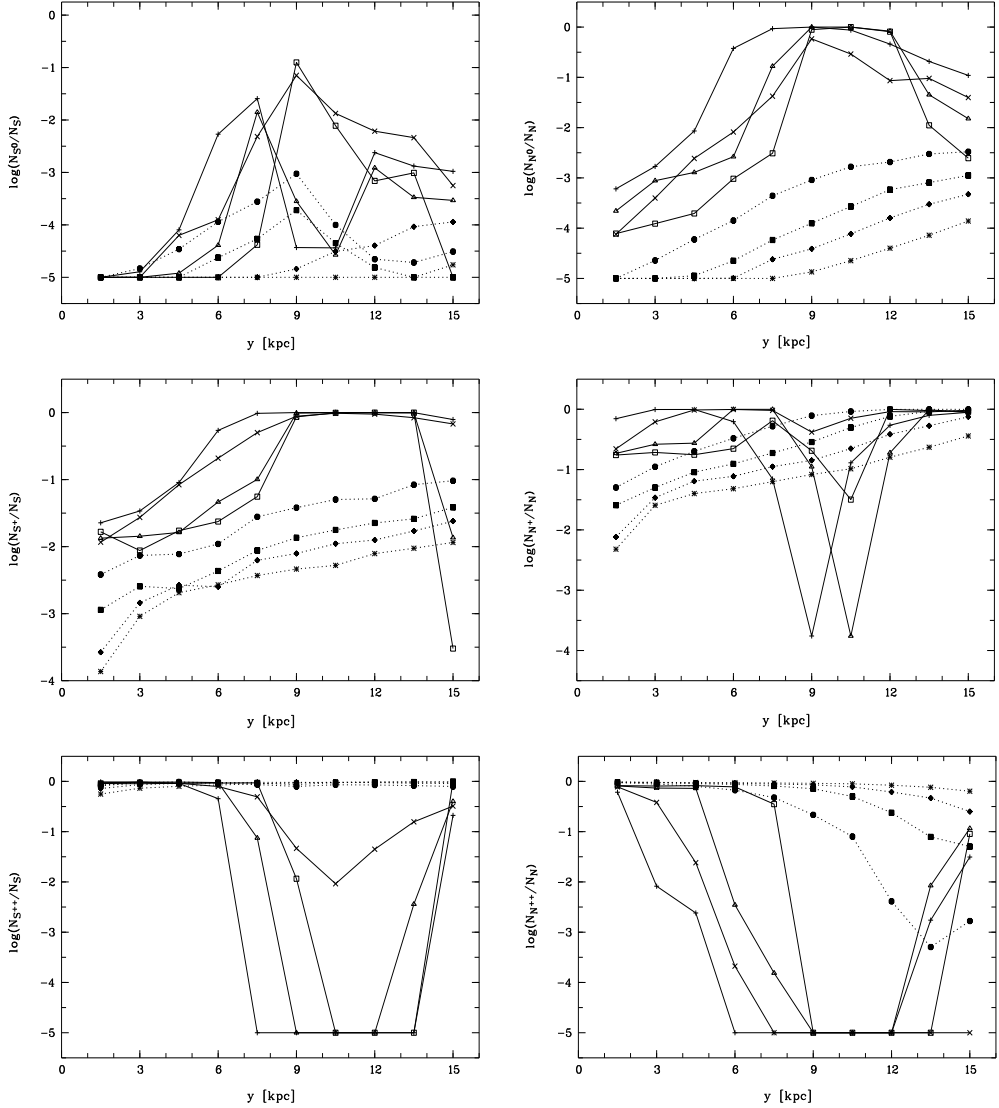}
\caption{Same as Fig.~\ref{f1}, except that fractions for sulfur (left panel) and nitrogen (right panel) are plotted.}
\label{f2}
\end{figure}

In Figures~\ref{f1} and \ref{f2} ionization fractions are plotted for the most important elements (H, He, O, N, and S) as a function of $y$ and $z$. For the sake of clarity, only data of 4 disk and 4 halo layers is shown, where each symbol/line addresses a certain distance above the galactic plane.  The ionization structures for H$^{0}$ and O$^{0}$ are nearly identical due to very similar ionization potentials. For $y\le9$\,kpc (radial direction) low fractions indicate that most of the hydrogen and oxygen is ionized. At distances $>9$\,kpc from the center the disk ends and the ionization structures change, implying that H and O are now almost completely neutral.  Similar effects are well known in normal HII-regions \citep[e.g.,][]{oster,och}.

The steep increase between 7 and 9\,kpc indicates that all photons are absorbed and cannot ionize the gas at larger $y$-distances.  This is the equivalent case to the Str\"omgren radius which determines the outer boundary of a classical HII-region. Between $11\,\mbox{kpc}<y<14$\,kpc there is a local minimum in both ionization fractions which demonstrates that H and O are partly ionized.  The reason is most likely due to ionizing photons which are created in the halo and are re-emitted at angles that allow them to reach the galactic plane. In other words, the diffuse radiation field is responsible for the ionization of the shadowed outer-disk region.

\begin{figure}[!t]
\plotfiddle{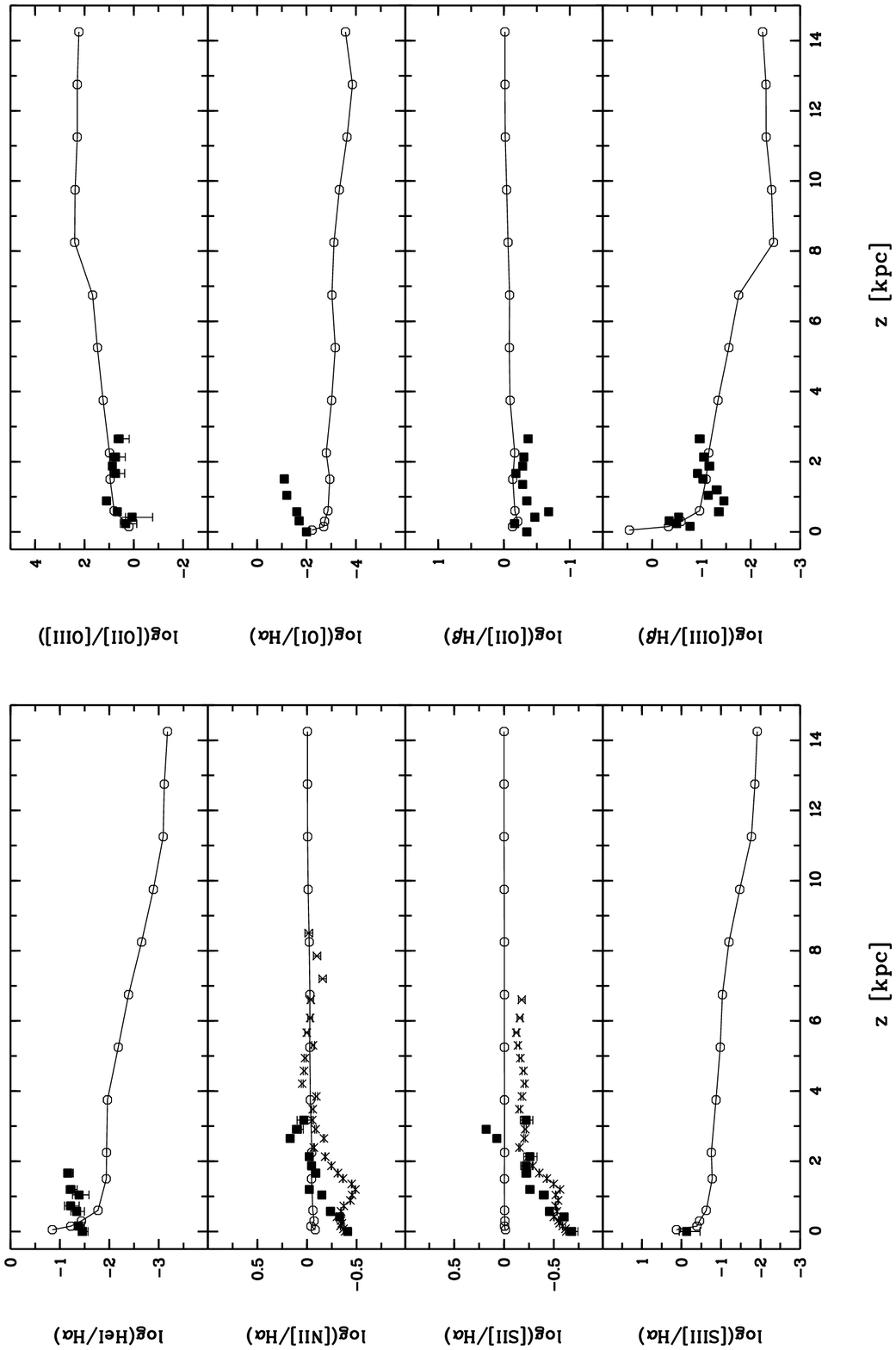}{7cm}{-90}{45}{45}{-190}{240}
\caption{Predicted line ratios for an inhomogeneous density halo. For a comparison with observations data of NGC\,891 \citep[][full squares]{otte02} is overplotted.  In addition, VLT-data of NGC\,5775 \citep[][``x''-signs]{tu00b} is also shown as it covers a substantial fraction of the model galaxy halo. 
Predictions for [SIII]$\lambda \lambda$9069,9531 are compared to the Mathis model \citep{ma00} since this line has not been observed in the DIG.}
\label{f3}
\end{figure}

If one moves along the minor axis the corresponding ionization fractions of $N_{H^{0}}/N_{H}$ and $N_{O^{0}}/N_{O}$ become more and more independent from the disk.  Beyond 9.0\,kpc the ratios drop continuously, implying that {\it with decreasing densities of the DIG, the fraction of ionized gas gets larger}. The corresponding curves for higher ionization stages of O, S, and N also support this finding. This surprising result is explainable as follows: stellar and diffuse ionizing photons that traversed the disk get dilute as they expand into the halo. Decreasing halo densities lead to decreasing opacities and enable photons to reach the outer halo border. This in turn implies low recombination rates and higher temperatures in the halo (cf. lower left panel of Fig.~\ref{f1}).  
Typical recombination times for hydrogen at gas temperatures of 8000\,K and densities of about $6\times10^{-4}$\,cm$^{-3}$ yield: $\tau_{rec}\approx1.2\times10^{8}$\,yr.
Apparently, the geometric dilution of the radiation field is not a severe obstruction for the photons to ionize nearly all atoms within a given cell volume.  Once the gas is ionized, it remains ionized for about 100 million years. If one considers the constant stream of ionizing photons from the disk it appears likely that recombination time scales are even longer.  Nevertheless, DIG seems to be far from ionization equilibrium. 

The same qualitative results and arguments also apply to heavier ions \citep[for a discussion see][]{tu02}. High ionization fractions of $N_{He^{+}}/N_{He}$ and $N_{N^{++}}/N_{N}$ close to 1 at large distances from the disk might suggest that the adopted radiation field produced by O6 stars is too hard.
Changing the spectral type of the ionizing sources to O7 or O8, would restrict the energy distribution function to less energetic photons and lead to a reduced amount of He$^{+}$ in the halo without substantially altering the ionization structure of the lower ionization species.

Line ratios presented in Fig.~\ref{f3} are normalized to H$\alpha$ or H$\beta$ and have been computed by summing up the volume averaged line intensities of all cells along the line of sight for each $z$-layer.
It turns out that our simple model is well suited to reproduce observed line ratios for NGC\,891, especially those of [OIII]/H$\beta$ and [OII]/H$\beta$. Deviations between predicted and observed ratios, as e.g., visualized in HeI$\lambda$5876/H$\alpha$ and [OI]$\lambda$6300/H$\alpha$, suggest that the adopted incident radiation field is too hard. However, pure photoionization by OB-stars seems to be appropriate to explain all observed line ratios consistently.

\begin{figure}
\plotfiddle{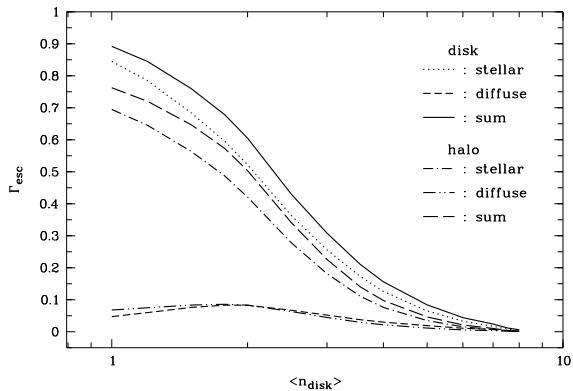}{5cm}{-90}{30}{30}{-130}{160}
\caption{$\Gamma_{esc}$, the fraction of stellar and diffuse ionizing photons that escaped the disk and the halo as a function of averaged disk densities $<\hspace{-0.1cm}n_{disk}\hspace{-0.1cm}>$.}
\label{f4}
\end{figure}

Finally, the photon statistics shown in Fig.~\ref{f4} allows to extract the artifical spectrum which leaves the disk and the halo. This plot also reveals a critical average density of about 8 $particles/cm^{3}$ at which no ionizing radiation escapes the galaxy. Below this value, a significant fraction of ionizing photons leave the halo and contributes to the intergalactic radiation field. A quantification of this fraction can be achieved after dust has been modeled in. Although estimates \citep{bh99,bh01} indicate that only a few percent of the ionizing photons contribute to the ionization of the IGM, this represents a significant contribution in view of the large number densities of normal spiral galaxies.   

\section{Summary and conclusions}
First results from a 3D self-consistent Monte Carlo model that accurately simulates DIG in galactic halos with respect to the geometry and the radiative transfer have been presented. The main conclusions can be summarized as follows:

(1) predicted line line ratios from a simple (dust-free) model agree reasonably well with those of NGC\,891, assuming O6 stars as ionizing sources and an exponentially decreasing gas density.

(2) all ionization fractions in the disk at $z\le0.2$\,kpc are in good agreement with predictions from model HII-regions \citep{och,oster}. We conclude that the disk can be approximated by a single huge HII-region.

(3) the degree of ionization seems to be higher in the halo than in the disk, as evidenced by the high fraction of ionized halo gas.

(4) apparently, the decreasing gas density leads to a larger fraction of ionized gas in the halo

(5) once the DIG is ionized, it remains ionized for typical time scales of about $10^8$\,yrs. 

(6) DIG seems to be in an extreme non-equilibrium condition.\\

Forthcoming work will involve a comprehensive parameter study in order to investigate the dependence between line ratios, gas densities, stellar photon luminosities, and gas phase abundances. After dusty plasmas can be simulated, detailed modeling of individual DIG properties of well observed galaxies, such as NGC\,5775 or NGC\,891, can be started.  In case theory and observations should reconcile, final proof would be given that photoionization by O stars can serve as the only ionization mechanism of the DIG if the physics of radiative transfer is calculated accurately and a realistic geometry is considered.

\end{document}